\documentclass[a4paper]{article}

\usepackage{geometry}
\usepackage{amsmath}
\usepackage{CJK}
\usepackage{chemarrow}
\usepackage{setspace}
\usepackage{array}
\usepackage{graphicx}
\usepackage{subfigure}
\usepackage{amsfonts}
\usepackage{amssymb}
\usepackage{latexsym}
\usepackage{accents}
 \usepackage{indentfirst}
\newtheorem{rem}{Remark}[section]

\newtheorem{Theorem}{Theorem}[section]

\newtheorem{prop}{Proposition}[section]

\begin{document}

\begin{spacing}{1.5}
\setlength{\parindent}{2em}
 \numberwithin{equation}{section}

\title{A rigorous derivation of multicomponent diffusion laws}
\author{Zaibao Yang\footnote{Zhou Pei-Yuan Center for Appl. Math., Tsinghua Univ., Beijing 100084, China; Email:
yang-zb11@mails.tsinghua.edu.cn},
~Wen-An Yong\footnote{Zhou Pei-Yuan Center for Appl. Math., Tsinghua Univ., Beijing 100084, China; Email:
Email: wayong@tsinghua.edu.cn},
~Yi Zhu\footnote{Zhou Pei-Yuan Center for Appl. Math., Tsinghua Univ., Beijing 100084, China; Email:
Email: yizhu@tsinghua.edu.cn}}
\date{}
\maketitle

\begin{abstract}
\boldmath
This article is concerned with the dynamics of a mixture of gases.
Under the assumption that all the gases are isothermal and inviscid, we show that the governing equations have an elegant conservation-dissipation structure. With the help of this structure, a multicomponent diffusion law is derived mathematically rigorously. This clarifies a long-standing non-uniqueness issue in the field for the first time. The multicomponent diffusion law derived here takes the spatial gradient of an entropic variable as the thermodynamic forces and satisfies a nonlinear version of the Onsager reciprocal relations.

\end{abstract}

\section{Introduction}
Multicomponent diffusion plays a dominant role in many industrial and natural processes and has been extensively studied in a large number of chemical engineering books and articles since 1948. See \cite{Fu,CH,Wi1,CC,Wi2,TK,CB,Gi,Lam,Da,MF, MR, RS, Ot1, VG} and the references cited therein. A fundamental task of these studies is to determine the relations between the thermodynamic fluxes and forces, which is called multicomponent diffusion laws or constitutive relations. In the literature, various different constitutive relations can be found and they are not always consistent with each other.
For example, in \cite{Lam} published in 2006 one can find the statement ``The fact that $D_{ij}$ (diffusion coefficients) is fundamentally nonunique is clearly documented in the literature.$^{1-4, 14-18}$". Therefore, there has been a urgent need for a mathematical clarification of the situation.

In this paper, we present a mathematically rigorous derivation of a multicomponent diffusion law for a mixture of gases and thereby clarify the uniqueness for the first time. For the sake of simplicity, we assume that all the gases are isothermal and inviscid. Our derivation follows the Furry-Williams approach \cite{Fu, Wi1, Lam, Da} starting from the macroscopic balance equations of all the gases and uses the Chapman-Enskog expansion or Maxwell iteration recently justified in \cite{YY} for a class of hyperbolic relaxation systems. The justification is crucially based on a conservation-dissipation structural property of the system of the macroscopic equations. The conservation-dissipation structure was firstly proposed and was shown in \cite{Yong2008} to be satisfied by many classical models from mathematical physics. Recently, it has been successfully used in \cite{ZHYY} to develop a general theory of mathematical modeling of irreversible processes.

The multicomponent diffusion law derived here takes the spatial gradient of an entropic variable as the thermodynamic forces and satisfies a nonlinear version of the Onsager reciprocal relations. Namely, the corresponding diffusion coefficient matrix is symmetric and positive-definite. The use of the entropic variable as thermodynamic forces is a key of the original Onsager relation \cite{On} and provides us a nice formulation of the multicomponent diffusion laws. This formulation is mathematically and numerically important in treating the multicomponent diffusion systems.

As mentioned earlier, there are various multicomponent diffusion laws in the literature and it is almost impossible to comment on all of them. For most of the existing multicomponent diffusion laws, the diffusion coefficient matrix is generally non-symmetric \cite{Da, MF} and sometimes even not completely determined \cite{Lam}. A further comparison of ours with that in \cite{Lam} is given in Section 4. On the other hand, the Maxwell iteration was also used in \cite{MR, RS} to derive the multicomponent diffusion laws. We also notice \cite{Ot1, VG} where the multicomponent diffusion laws were derived in two different frameworks of non-equilibrium thermodynamics. Unfortunately, all these derivations are formal, lack a mathematically rigorous justification, and therefore the uniqueness issue has been not resolved before.

The paper is organized as follows. In Section 2, we introduce the classical mass and momentum equations for multicomponent diffusion systems. Section 3 focuses on the conservation-dissipation structure of the systems. The multicomponent diffusion law is derived in Section 4.

\section{Governing equations}
Consider a mixture of $N$ gases. Assume that all the gases are inviscid and isothermal. Then the motion of the gases obeys the classical mass and momentum equations \cite{Lam, Da}.
Let $\rho_i$ and $\mathbf{V}_i$ denote the density and mass averaged velocity of species $i$, respectively. The species-specific mass equations read as
\begin{equation}\label{21}
\dfrac{\partial \rho_i}{\partial t}+\nabla\cdot (\rho_i\mathbf{V}_i)=\dot {W}_i, \quad i=1,\cdots, N,
\end{equation}
where $\dot {W}_i$ represents the net mass production rate (per unit volume) of species  $i$ due to chemical reactions. Since mass can either be created or destroyed by chemical reactions, the $\dot W_i$'s always satisfy
\begin{equation*}
\sum\limits_{i=1}^N\dot W_i=0.
\end{equation*}

On the other hand, the momentum equation for species $i$ is
\begin{equation}\label{22}
\dfrac{\partial(\rho_i\mathbf{V}_i)}{\partial t}+\nabla\cdot(\rho_i\mathbf{V}_i\mathbf{V}_i+p_iI_d)=\rho_i\mathbf {f}_i+\dot W_i\mathbf{V}_i+\mathbf G_i^{coll},\quad i=1,\cdots,N.
\end{equation}
Here $p_i\equiv p_i(\rho_i)$ is the pressure of species $i$, $I_d$ is the unit matrix of order $d$ ($d=1, 2, 3$), $\rho_i\mathbf{f}_i$ is the body force acting on species  $i$, and $\mathbf G_i^{coll}$ stands for the net impacts of all interspecies collisional momentum exchanges on species  $i$.
Since interspecies collisions conserve the total momentum, the sum of $\mathbf{G}_i^{coll}$ over all species must be identically zero:
\begin{equation}\label{23}
\sum\limits_{i=1}^N\mathbf G_i^{coll}=0.
\end{equation}

By its definition, $\mathbf G_i^{coll}$ is a function of $\mathbf{X}=(\mathbf{V}_1-\mathbf{V}_i,\cdots,\mathbf{V}_N-\mathbf{V}_i)$ and vanishes whenever $\mathbf{X}=\mathbf{0}$. The latter means that there is no collisional momentum exchange with species $i$ when other species all have the same velocity.
Based on this observation, we may write
 $$
 \mathbf{G}_i^{coll}=\int_0^1\frac{\partial \mathbf{G}_i^{coll}}{\partial \mathbf{X}}(\theta \mathbf X)d\theta \cdot \mathbf X\equiv \sum\limits_{j\ne i} K_{ij}(\mathbf V_i-\mathbf V_j)
 $$
 with $K_{ij}=[K_{ij}^{\alpha\beta}]_{d\times d}$ being $d\times d$-matrixes. $K_{ij}^{\alpha\beta}$ is called a collisional coefficient, possibly depending on any quantities like densities under consideration.

For the Stefan-Maxwell model \cite{Fu, Wi1, Lam}, each $K_{ij}$ is a scalar matrix, that is,
$$
K_{ij}^{\alpha\beta}=-\sigma_{ij}\delta_{\alpha\beta},\qquad \sigma_{ij}=\bar m_{ij}\nu_{ij}.
$$
Here $\delta_{\alpha\beta}$ is the Kronecker delta, $\bar m_{ij}=m_im_j/(m_i+m_j)$ is the ``reduced mass", $m_i$ is the molecular mass of species $i$, and $\nu_{ij}$ is the
averaged frequency (per unit volume) of collisions between molecules $i$ and $j$, and satisfies
$$
\sigma_{ij}=\sigma_{ji}>0
$$
for each $i\neq j$ and $i,j=1,\cdots, N$.

For future references, we rewrite $\mathbf G_i^{coll}$ as
$$
\mathbf G_i^{coll}= -\sum\limits_{j=1}^NK_{ij}\mathbf{V}_j
$$
with
$$
K_{ij}=\delta_{ij}\sum\limits_{k=1}^N\sigma_{ik}-\sigma_{ij}
$$
for each $i,j=1,\cdots,N$. Obviously, $K_{ij}\equiv K_{ij}(\rho_1,\rho_1\mathbf V_1,\cdots,\rho_N,\rho_N\mathbf V_N)$ satisfies
\begin{equation}\label{24}
\sum\limits_{l=1}^NK_{il}\equiv 0\qquad \text{and} \qquad
K_{ij}=K_{ji}<0, \quad  i\neq j,
\end{equation}
for $i,j=1,\cdots,N$. This is consistent with \eqref{23}.

In a mixture of gases, collisions usually happen much faster than the macroscopic fluid motions. Denote by $\varepsilon$ the ratio of characteristic collision time to characteristic fluid mechanics time. Then $\varepsilon$ is small and the collision coefficient $K_{ij}$ may be scaled as
\begin{equation}\label{25}
K_{ij}=\dfrac{1}{\varepsilon}{\tilde K}_{ij}
\end{equation}
with $\tilde K_{ij}=O(1)$.
For the notational convenience, we will use $K_{ij}$ for $\tilde K_{ij}$ in the rest of this paper.

\section{Conservation-dissipation structure}
In this section, we show that the system of equations \eqref{21} and \eqref{22} with fast collisions \eqref{25} possesses the conservation-dissipation structure proposed in \cite{Yong2008}.
To do this, we rewrite \eqref{21} and \eqref{22} as
\begin{equation}\label{31}
\begin{split}
&\dfrac{\partial \mathbf{U}_i}{\partial t}+\sum\limits_{j=1}^d\dfrac{\partial\mathbf{\hat F}_j(\mathbf{U}_i)}{\partial x_j}=\dfrac{1}{\varepsilon}Q_i(\mathbf{U}),\quad i=1,2,\cdots,N\\
\text{or}\qquad \qquad &\\
&\dfrac{\partial \mathbf{U}}{\partial t}+\sum\limits_{j=1}^d\dfrac{\partial\mathbf{F}_j(\mathbf{U})}{\partial x_j}=\dfrac{1}{\varepsilon}Q(\mathbf{U})
\end{split}
\end{equation}
with the external forces $\rho_i\mathbf{f}_i$ and chemical reaction terms $\dot{W}$ being ignored to simplify the exposition.
In \eqref{31}, $d$ is the space dimension,
$$
\mathbf U_i=\left( \begin{array}{c}
 {\rho _i} \\
 {\rho _i}{\mathbf V_i} \\
 \end{array} \right), \qquad  \mathbf{ \hat F}_j(\mathbf U_i)=\left( \begin{array}{c}
 {\rho _i V_{i,j}} \\
 {\rho _i}{\mathbf V_{i}V_{i,j}}+p_i\mathbf e_j \\
 \end{array} \right),\qquad Q_i(\mathbf U)=\left( \begin{array}{c}
 {0} \\
 {-\sum\limits_{k=1}^N K_{ik}\mathbf V_k} \\
 \end{array} \right),
 $$
$\mathbf{U}=(\mathbf{U}_1,\mathbf{U}_2,\cdots,\mathbf{U}_N)^T, \quad \mathbf{F}_j(\mathbf{U})=(\mathbf{\hat F}_j(\mathbf{U}_1),\cdots,\mathbf{\hat F}_j(\mathbf{U}_N))^T,\quad Q(\mathbf{U})=(Q_1(\mathbf{U}),\cdots,Q_N(\mathbf{U}))^T$,
$V_{i,j}$ denotes the $j$-th component of velocity $\mathbf V_i$, and $\mathbf e_j$ stands for the $j$-th column of the unit matrix $I_d$.

Following \cite{Yong2014}, we define a mathematical entropy $\eta_i$ for species $i$:
$$
\eta_i(\rho_i,\rho_i\mathbf{V}_i)=\rho_i\int_{\bar\rho_i}^{\rho_i}\dfrac{p_i(z)}{z^2}dz
+\dfrac{1}{2\rho_i}|\rho_i\mathbf{V}_i|^2
$$
with $\bar\rho_i$ a possible positive value.
Consequently, a mathematical entropy $\eta(\mathbf U)$ for the mixture can be introduced as
\begin{equation}\label{32}
\eta(\mathbf U)=\sum\limits_{i=1}^N\rho_i\int_{\bar\rho_i}^{\rho_i}\dfrac{p_i(z)}{z^2}dz+\sum\limits_{i=1}^N\dfrac{1}{2\rho_i}|\rho_i\mathbf{V}_i|^2.
\end{equation}

The conservation-dissipation structure for system \eqref{31} can be stated as follows.
\begin{Theorem}
Assume that \eqref{24} holds true and $p_i=p_i(\rho_i)$ is strictly increasing for $\rho_i>0$ and for each $i=1,\cdots,N$. Then the mathematical entropy $\eta(\mathbf U)$ defined in \eqref{32} for system \eqref{31} is strictly convex in $O_U=\{\mathbf U\in R^{dN}\ | \ \rho_i>0, i=1,\cdots,N\}$.  Moreover, the following three statements are true:
\begin{itemize}
  \item $\dfrac{\partial^2\eta(\mathbf{U})}{\partial\mathbf{U}^2}\dfrac{\partial\mathbf{F}_{j}(\mathbf{U})}{\partial\mathbf{U}}$ is symmetric for each $\mathbf U\in O_U$ and for each $j$;
  \item There exists a symmetric and semi-positive matrix $L(\mathbf U)$ such that
  $$Q(\mathbf U)=-L(\mathbf U)\dfrac{\partial\eta(\mathbf U)}{\partial\mathbf U};$$
  \item The null-space of $L(\mathbf U)$ is independent of $\mathbf U\in O_U$.
\end{itemize}
\end{Theorem}

\begin{rem}
The first statement is the well-known entropy condition for hyperbolic conservation laws and corresponds
to the classical principles of thermodynamics. The second one can be understand as a nonlinearization of the
celebrated Onsager reciprocal relation in modern thermodynamics \cite{Groot84} and implies the second law
of thermodynamics. It displays a direct relation of irreversible processes to the entropy change. The last one expresses the fact that physical laws of conservation hold true, no matter what state
the underlying thermodynamical system is in (equilibrium, non-equilibrium, and so on).
\end{rem}

Here is our proof of the above theorem. Compute
$$
\dfrac{\partial\eta(\mathbf{U})}{\partial\mathbf U_i}=\left(\begin{array}{c}
{\dfrac{\partial \eta_i}{\partial \rho_i}}\\[4mm]
{\mathbf V_i}
\end{array} \right), \quad \dfrac{\partial^2\eta(\mathbf{U})}{\partial\mathbf U_i\partial\mathbf U_j}=\dfrac{1}{\rho_i}\left[ {\begin{array}{*{20}{c}}
   p'_i(\rho_i)+|\mathbf V_i|^2 & -\mathbf V_i^T  \\[4mm]
   -\mathbf V_i & I_d  \\
\end{array}} \right]\delta_{ij}.
$$
Because $p_i'(\rho_i)>0$ for $\rho_i>0$, one can directly verify that the Hessian $\dfrac{\partial^2\eta(\mathbf{U})}{\partial\mathbf U^2}$ is positive definite for each $\mathbf U\in O_U$. Therefore, $\eta=\eta(\mathbf{U})$ is strictly convex in $O_U$.

Next we compute from the definition of $\hat{\mathbf{F}}_j(\mathbf U_i)$ that, for $i\ne k$,
$$
\dfrac{\partial^2 \eta(\mathbf U)}{\partial \mathbf U_i\partial \mathbf U_k}=0, \qquad \dfrac{\partial \hat{\mathbf F}_j(\mathbf U_i)}{\partial \mathbf U_k}=0,
$$
and
$$
\dfrac{\partial^2\eta(\mathbf{U})}{\partial\mathbf U_i^2}\dfrac{\partial \hat{\mathbf F}_j(\mathbf U_i)}{\partial \mathbf U_i}=\dfrac{1}{\rho_i}\left[{\begin{array}{*{20}{c}} V_{i,j}|\mathbf V_i|^2 - p_i'(\rho_i){\mathbf V}_i^T{\mathbf e}_j& p'_i(\rho_i)\mathbf e_j^T-V_{i,j}\mathbf V_i^T\\[4mm]
p'_i(\rho_i)\mathbf e_j-V_{i,j}\mathbf V_i & V_{i,j}I_d\\
\end{array}}\right].
$$
Therefore, $\dfrac{\partial^2\eta(\mathbf{U})}{\partial\mathbf{U}^2}\dfrac{\partial\mathbf{F}_{j}(\mathbf{U})}{\partial\mathbf{U}}$ is symmetric.

Now we recall the structure of $Q(\mathbf{U})$ and define a $(d+1)N\times (d+1)N$ matrix $L(\mathbf{U})\equiv [L_{ik}(\mathbf U)]_{N\times N}$ with
$$
 L_{ik}(\mathbf U)=\left[{\begin{array}{*{20}{c}} 0 & 0\\
0 & K_{ik}I_d\\
\end{array}}\right].
$$
Then $Q(\mathbf U)$ can be written as
$$
Q(\mathbf U)=-L(\mathbf U)\dfrac{\partial\eta(\mathbf U)}{\partial\mathbf U}.
$$
From \eqref{24} it is not difficult to see that $L(\mathbf U)$ is symmetric and nonnegative-definite matrix.

It remains to show that the null-space of $L(\mathbf{U})$ is independent of $\mathbf U$. To do this, we write $\mathbf w\in R^{(d+1)N}$ as $\mathbf w=\left( \begin{array}{l}
 \mathbf w_1 \\
  \vdots  \\
 \mathbf w_N \\
 \end{array} \right)$
 with $\mathbf w_k=\left( \begin{array}{l}
 w_k^I \\
 \mathbf w_k^{II} \\
 \end{array} \right)$ and $\mathbf w_k^{II}\in R^d$.
If $L(\mathbf U)\cdot \mathbf w=0$, then
$$\sum\limits_{k=1}^NL_{ik}\mathbf {w}_k=\sum\limits_{k=1}^N\left[{\begin{array}{*{20}{c}} 0 & 0\\
0 & K_{ik}I_d\\
\end{array}}\right]\left( \begin{array}{l}
 {w_k ^I} \\
 {\mathbf {w}_k ^{II}} \\
 \end{array} \right)=\left( \begin{array}{c}
 {0} \\
 {\sum\limits_{k=1}^NK_{ik}\mathbf{w}_k^{II}} \\
 \end{array} \right)=0$$
 for each $i$. Since
 the null-space of the $N\times N$-matrix $K=[K_{ij}]$ is spanned by \{$(1,\cdots,1)^T$\}, it follows that
 $$
\mathbf{w}_1^{II}=\mathbf{w}_2^{II}=\cdots=\mathbf{w}_N^{II} .
 $$
Therefore, the null-space of $L(\mathbf U)$ is
$$
\Big\{\mathbf w\in R^{(d+1)N}: (\mathbf{w}_1^{I}, \mathbf{w}_2^{I}, \cdots, \mathbf{w}_N^{I})\in R^{N} \quad \mbox{and} \quad \mathbf{w}_1^{II}=\mathbf{w}_2^{II}=\cdots=\mathbf{w}_N^{II}\in R^{d} \Big\},
$$
which is independent of $U$. This completes the proof.

\section{Multicomponent diffusion laws}

In this section, we derive a multicomponent diffusion law by using the Maxwell iteration \cite{MR, RS} or Chapman-Enskog expansion justified in our recent paper \cite{YY}. The conservation-dissipation structure provided a highly efficient framework for the justification.

To begin with, we introduce the mass diffusion flux (also called thermodynamic flux)
$$
\mathbf{J}_i\equiv \rho_i(\mathbf{V}_i-\mathbf{V}),
$$
where the mass-averaged velocity $\mathbf{V}$ is defined as
$$
\mathbf{V}\equiv \dfrac{1}{\rho}\sum\limits_{i=1}^N\rho_i\mathbf{V}_i,\qquad \rho\equiv \sum\limits_{i=1}^N\rho_i.
$$
From these definitions, it follows immediately that
$$
\sum\limits_{i=1}^N\mathbf{J}_i=0.
$$
This is called zero-net-flux condition in some literature \cite{Lam}. Due to this condition, we only need to  consider the first $(N-1)$ fluxes $\mathbf J_i$ ($i=1,\cdots, N-1$).

Referring to \cite{YY}, we rewrite the multicomponent diffusion system \eqref{31} in term of the new variable
$$
\mathbf W\equiv (\rho,\rho \mathbf V, \rho_1,\cdots,\rho_{N-1},\mathbf J_1,\cdots,\mathbf J_{N-1})^T
\longleftarrow
\mathbf U\equiv(\rho_1,\rho_1\mathbf V_1,\cdots,\rho_N,\rho_N\mathbf V_N)^T.
$$
Firstly, the $N$ mass equations are equivalent to
\begin{equation}\label{41}
\begin{split}
&\dfrac{\partial \rho}{\partial t}+\nabla \cdot (\rho \mathbf V)=0,\\
&\dfrac{\partial \rho_i}{\partial t}+\nabla\cdot (\rho_i\mathbf V + \mathbf J_i)=0, \qquad i=1,\cdots,N-1,
\end{split}
\end{equation}
which is derived by summing up the mass equations \eqref{21} over $i$ and using the definition $\mathbf J_i=\rho_i(\mathbf V_i-\mathbf V)$. Recall that the external forces and reaction sources have been neglected. In order to obtain the equations for the total momentum $\rho \mathbf V$ and the first $(N-1)$ fluxes $\mathbf J_i$, we recast the left-hand side of the momentum equation \eqref{22} as
\begin{equation*}
\begin{split}
&\dfrac{\partial (\rho_i\mathbf V_i)}{\partial t}+\nabla\cdot(\rho_i\mathbf V_i\otimes\mathbf V_i)+\nabla p_i \\
&=\dfrac{\partial \mathbf J_i}{\partial t} + \dfrac{\partial (\rho_i\mathbf V)}{\partial t}+\nabla\cdot(\rho_i\mathbf V\otimes\mathbf V)+\nabla p_i+\nabla\cdot (\mathbf J_i\otimes\mathbf V+\mathbf V\otimes\mathbf J_i+\dfrac{\mathbf J_i\otimes \mathbf J_i}{\rho_i}).
\end{split}
\end{equation*}
Summing up these momentum equations over $i$, we use \eqref{23} and the zero-net-flux condition to obtain
\begin{equation}\label{42}
\begin{split}
&\dfrac{\partial (\rho\mathbf V)}{\partial t}+\nabla \cdot(\rho \mathbf V \otimes \mathbf V)+\nabla\sum\limits_{j=1}^Np_j+ \sum\limits_{j=1}^N \nabla \cdot(\frac{\mathbf J_j \otimes \mathbf J_j}{\rho_j})=0.
\end{split}
\end{equation}
Moreover, we use $\sum\limits_{j=1}^NK_{ij} =0$ in \eqref{24} and rewrite the collision term
\begin{equation*}
\begin{split}
\mathbf G_i^{coll}&=-\dfrac{1}{\varepsilon}\sum\limits_{j=1}^NK_{ij}\mathbf ({\mathbf V_j}-{\mathbf V})\\
&=-\dfrac{1}{\varepsilon}\Big(\sum\limits_{j=1}^{N-1}K_{ij}\dfrac{\mathbf J_j}{\rho_j}+K_{iN}\dfrac{\mathbf J_N}{\rho_N}\Big)\\
&=-\dfrac{1}{\varepsilon}\sum\limits_{j=1}^{N-1}K_{ij}(\dfrac{\mathbf J_j}{\rho_j}-\dfrac{\mathbf J_N}{\rho_N})\\
&=-\dfrac{1}{\varepsilon}\sum\limits_{j=1}^{N-1}K_{ij}(\dfrac{\mathbf J_j}{\rho_j}+\dfrac{1}{\rho_N}\sum\limits_{l=1}^{N-1}\mathbf J_l)\\
&=-\dfrac{1}{\varepsilon}\sum\limits_{j,l=1}^{N-1}K_{ij}\big{(}\dfrac{1}{\rho_l}\delta_{jl}+\dfrac{1}{\rho_N}\big{)}\mathbf J_l.
\end{split}
\end{equation*}
Thus, the equation for $\mathbf J_i$ reads as
\begin{equation}\label{43}
\begin{split}
&\dfrac{\partial \mathbf J_i}{\partial t} +\dfrac{\partial (\rho_i\mathbf V)}{\partial t}+\nabla\cdot(\rho_i\mathbf V\otimes\mathbf V)+\nabla p_i+\nabla\cdot \big{(}\mathbf V\otimes \mathbf J_i+\mathbf J_i\otimes \mathbf V+\dfrac{\mathbf J_i\otimes \mathbf J_i}{\rho_i} \big{)} \\ &=-\dfrac{1}{\varepsilon}\sum\limits_{j,l=1}^{N-1}K_{ij}\big{(}\dfrac{1}{\rho_l}\delta_{jl}+\dfrac{1}{\rho_N}\big{)}\mathbf J_l.
\end{split}
\end{equation}
Consequently, system \eqref{31} has been rewritten as \eqref{41}--\eqref{43}.

On the other hand, we recall \eqref{24} that $K_{ij}=K_{ji}<0$ with $i\neq j$ and $\sum\limits_{j=1}^NK_{ij}=0$. Then  $[K_{ij}]_{(N-1)\times(N-1)}$ is a strictly diagonally dominant and symmetric matrix. Therefore, it has an inverse, say $\mathcal K$,  and the inverse is symmetric and positive definite. Moreover, we set \cite{RS}
$$
\Phi_{ij}=\dfrac{1}{\rho_j}\delta_{ij}+\dfrac{1}{\rho_N}
\quad \mbox{and} \quad C_{ij}=\rho_j\delta_{ij}-\dfrac{\rho_i\rho_j}{\rho}
$$
for $i,j=1,\cdots,N-1$. Note that
\begin{equation*}
\begin{split}
(C\Phi)_{ij}&=\sum\limits_{l=1}^{N-1}C_{il}\Phi_{lj}=\sum\limits_{l=1}^{N-1}(\rho_l\delta_{il}-\dfrac{\rho_i\rho_l}{\rho})(\dfrac{1}{\rho_j}\delta_{lj}+\dfrac{1}{\rho_N})\\
&=\delta_{ij}+\dfrac{\rho_i}{\rho_N}-\dfrac{\rho_i}{\rho}-\dfrac{\rho_i}{\rho\rho_N}\sum\limits_{l=1}^{N-1}\rho_l\\
&=\delta_{ij}+\dfrac{\rho_i}{\rho_N}-\dfrac{\rho_i}{\rho}-\dfrac{\rho_i}{\rho\rho_N}(\rho-\rho_N)\\
&=\delta_{ij}.
\end{split}
\end{equation*}
and
\begin{equation*}
\begin{split}
\dfrac{\partial (\rho_i\mathbf V)}{\partial t}+\nabla\cdot(\rho_i\mathbf V\otimes\mathbf V)=& \rho_i(\mathbf V_t+\mathbf V\cdot \nabla \mathbf V) + +(\rho_{it}+\nabla\cdot(\rho_i\mathbf V))\mathbf V\\
=& \rho_i(\mathbf V_t+\mathbf V\cdot \nabla \mathbf V)-(\nabla\cdot \mathbf J_i)\mathbf V
\end{split}
\end{equation*}
due to the second line in \eqref{41}. The flux equation \eqref{43} can be rewritten as
\begin{equation}\label{44}
\begin{split}
\mathbf J_i& =-\varepsilon\sum\limits_{k,l=1}^{N-1}C_{ik}\mathcal{K}_{kl}\Big{(}\rho_l(\mathbf V_t+\mathbf V\cdot \nabla \mathbf V)+\nabla p_l -(\nabla\cdot \mathbf J_l)\mathbf V +(\mathbf J_l)_t+\nabla\cdot \big{(}\mathbf V\otimes \mathbf J_l+\mathbf J_l\otimes \mathbf V+\dfrac{\mathbf J_l\otimes \mathbf J_l}{\rho_l} \big{)}\Big{)}\\
&=-\varepsilon\sum\limits_{k,l=1}^{N-1}C_{ik}\mathcal{K}_{kl}\Big{(}-\dfrac{\rho_l}{\rho}\nabla\sum\limits_{j=1}^Np_j+\nabla p_l+O(\varepsilon) \Big{)}.
\end{split}
\end{equation}
The second step is due to \eqref{42} and the Maxwell iteration, using that $\mathbf J_i=O(\varepsilon)$ indicated by the first step.

Furthermore, we deduce from \eqref{44} that
\begin{equation*}
\begin{split}
\mathbf J_i
&=-\varepsilon\sum\limits_{k,l=1}^{N-1}C_{ik}\mathcal{K}_{kl}\Big{(}-\sum\limits_{j=1}^{N-1}\dfrac{\rho_j\rho_l}{\rho}\dfrac{\nabla p_j}{\rho_j}-\dfrac{\rho_N\rho_l}{\rho}\dfrac{\nabla p_N}{\rho_N}+\rho_l\dfrac{\nabla p_l}{\rho_l} \Big{)}+O(\varepsilon^2)\\
&=-\varepsilon\sum\limits_{k,l=1}^{N-1}C_{ik}\mathcal{K}_{kl}\Big{(}\sum\limits_{j=1}^{N-1}(\rho_j\delta_{lj}-\dfrac{\rho_j\rho_l}{\rho})\dfrac{\nabla p_j}{\rho_j}-\dfrac{\rho_l}{\rho}(\rho-\sum\limits_{j=1}^{N-1}\rho_j)\dfrac{\nabla p_N}{\rho_N} \Big{)}+O(\varepsilon^2)\\
&=-\varepsilon\sum\limits_{k,l=1}^{N-1}C_{ik}\mathcal{K}_{kl}\Big{(}\sum\limits_{j=1}^{N-1}(\rho_j\delta_{lj}-\dfrac{\rho_j\rho_l}{\rho})\dfrac{\nabla p_j}{\rho_j}-\sum\limits_{j=1}^{N-1}(\rho_j\delta_{lj}-\dfrac{\rho_j\rho_l}{\rho})\dfrac{\nabla p_N}{\rho_N} \Big{)}+O(\varepsilon^2)\\
&=-\varepsilon\sum\limits_{k,l,j=1}^{N-1}C_{ik}\mathcal{K}_{kl}(\rho_j\delta_{lj}-\dfrac{\rho_j\rho_l}{\rho})(\dfrac{\nabla p_j}{\rho_j}-\dfrac{\nabla p_N}{\rho_N})+O(\varepsilon^2)\\
&=-\varepsilon\sum\limits_{k,l,j=1}^{N-1}C_{ik}\mathcal{\bar K}_{kl}C_{lj}\big{(}\dfrac{\nabla p_j}{\rho_j}-\dfrac{\nabla p_N}{\rho_N}\big{)}+O(\varepsilon^2)
\end{split}
\end{equation*}
for $i=1,\cdots, N-1$. In the last step we have approximated $\mathcal{K}=\mathcal{K}(\rho, \rho\mathbf V, \rho_1,\cdots,\rho_{N-1}, \mathbf J_1, \cdots, \mathbf J_{N-1})$ with $\mathcal{\bar K}=\mathcal{K}(\rho, \rho\mathbf V, \rho_1,\cdots,\rho_{N-1}, 0, \cdots, 0)$, that is, the latter is evaluated at the equilibrium.
Truncating the expansion above, we arrive at the multicomponent diffusion law
\begin{equation}\label{45}
\mathbf J_i=-\varepsilon\sum\limits_{j=1}^{N-1}D_{ij}(u)\big{(}\dfrac{\nabla p_j}{\rho_j}-\dfrac{\nabla p_N}{\rho_N}\big{)}
\end{equation}
with $D_{ij}(u)\equiv \sum\limits_{k,l=1}^{N-1}C_{ik}\mathcal{\bar K}_{kl}C_{lj}$ the multicomponent diffusion coefficients, which depend only on the conserved variable $u\equiv (\rho, \rho\mathbf V, \rho_1,\cdots,\rho_{N-1})^T$.

As to the relation \eqref{45}, we have the following remarks.

\begin{rem}
\begin{itemize}

\item The diffusion matrix
$$
D(u)\equiv [D_{ij}(u)]_{(N-1)\times(N-1)}=[C_{ij}]_{(N-1)\times(N-1)}\mathcal {\bar K}[C_{ij}]_{(N-1)\times(N-1)}
$$
is symmetric and positive definite, since $[C_{ij}]_{(N-1)\times(N-1)}$ is invertible, $\mathcal {\bar K}$ is positive definite and they are both symmetric.

\item According to the general theory (see, e.g., \cite{YY}), the term $\big{(}\dfrac{\nabla p_j}{\rho_j}-\dfrac{\nabla p_N}{\rho_N}\big{)}$ can be expressed as the spatial gradient of the equilibrium-entropic force. To see this, we recall \eqref{32} that
\begin{equation*}
\begin{split}
\eta(\mathbf U)=&\sum\limits_{i=1}^N\Big(\rho_i\int_1^{\rho_i}\dfrac{p_i(z)}{z^2}dz + \dfrac{1}{2\rho_i}|\rho_i\mathbf{V}_i|^2\Big)\\
=& \sum\limits_{i=1}^N\Big(\rho_i\int_1^{\rho_i}\dfrac{p_i(z)}{z^2}dz + \dfrac{1}{2\rho_i}|\rho_i\mathbf{V} + \mathbf J_i|^2\Big).
\end{split}
\end{equation*}
Thus, we have
\begin{equation*}
\eta^{eq}(u)\equiv\eta(\mathbf U)|_{\mathbf J_i=0}=\sum\limits_{i=1}^N\rho_i\int_1^{\rho_i}\dfrac{p_i(z)}{z^2}dz + \dfrac{\rho}{2}|\mathbf{V}|^2.
\end{equation*}
For fixed $\rho$ and $\mathbf{V}$, we compute
\begin{equation*}
\nabla \dfrac{\partial\eta^{eq}(u)}{\partial \rho_i}
=\dfrac{\nabla p_i}{\rho_i}-\dfrac{\nabla p_N}{\rho_N}, \quad i=1, 2, \cdots, N-1.
\end{equation*}
Consequently, the multicomponent diffusion law \eqref{45} can be rewritten as
$$
\mathbf J_i=-\varepsilon\sum\limits_{j=1}^{N-1}D_{ij}(u)\nabla \dfrac{\partial\eta^{eq}(u)}{\partial \rho_j}.
$$
Namely, the thermodynamic fluxes $\mathbf J_i$ are expressed in term of the equilibrium-entropic (thermodynamic) forces $\nabla \dfrac{\partial\eta^{eq}(u)}{\partial \rho_j}$, with the coefficient matrix $D(u)$ symmetric. This looks like the Onsager reciprocal relation \cite{On}, while the definition of the thermodynamic forces is consistent to those in the literature \cite{MR,RS}.

\end{itemize}
\end{rem}

With the diffusion law \eqref{45}, the system \eqref{41}--\eqref{43} can be approximated formally by the following second-order partial differential equations
\begin{equation*}
\begin{split}
&\dfrac{\partial \rho}{\partial t}+\nabla \cdot (\rho \mathbf V)=0,\\
&\dfrac{\partial (\rho\mathbf V)}{\partial t}+\nabla \cdot(\rho \mathbf V \otimes \mathbf V)+\nabla\sum\limits_{j=1}^Np_j=0,\\
&\dfrac{\partial \rho_i}{\partial t}+\nabla\cdot (\rho_i\mathbf V)=\varepsilon\nabla\cdot \sum\limits_{j=1}^{N-1}D_{ij}(u)(\dfrac{\nabla p_j}{\rho_j}-\dfrac{\nabla p_N}{\rho_N})
\end{split}
\end{equation*}
for $i=1,\cdots,N-1$. Thanks to the remark above, the last equations can be written as
\begin{equation}\label{46}
\begin{split}
u_t+\nabla\cdot G(u)=\varepsilon \left( \begin{array}{c}
 0 \\
\sum\limits_{k=1}^d \left( {D(u)(\frac{{\partial {\eta ^{eq}}(u)}}{{\partial \nu }})_{x_k}} \right)_{x_k} \\
 \end{array} \right)
\end{split}
\end{equation}
with $G(u)=(\rho\mathbf V,\rho\mathbf V\otimes\mathbf V+\sum\limits_{j=1}^Np_jI_d,\rho_1\mathbf V,\cdots,\rho_{N-1}\mathbf V)^T$ and $\nu=(\rho_1,\cdots,\rho_{N-1})^T$. This is the second-order partial differential equations (2.5) in \cite{YY} with
$$
B^{jk}(u) = \delta_{jk}\mbox{diag}(0_{d+1}, D(u))\eta^{eq}_{uu}(u),
$$
where $0_{d+1}$ is the zero-matrix of order $(d+1)$. It is known from \cite{YY} that $\eta^{eq}(u)$ is strictly convex and therefore its Hessian $\eta^{eq}_{uu}(u)$ is positive definite. Moreover, it was showed in \cite{YY} that  the system \eqref{46} has a nice entropy structure, which is extremely important mathematically and numerically.

With the above expression of $B^{jk}(u)$, we can simply show the following proposition.

\begin{prop}\label{P}
The system \eqref{31} satisfies the isotropy condition (*) of Theorem 2.1 in \cite{YY}.
\end{prop}
{\bf Proof}. By Lemma 3.2 in \cite{YY}, it suffices to show that the null-space of the symbol matrix
$$B(u,\xi)\equiv \sum\limits_{j,k=1}^d B^{jk}(u)\big(\eta^{eq}_{uu}(u)\big)^{-1}\xi_j\xi_k
=\sum\limits_{j,k=1}^d \delta_{jk}\mbox{diag}(0_{d+1}, D(u))\xi_j\xi_k
=\mbox{diag}(0_{d+1}, D(u))|\xi|^2
$$
is independent of $u$ and $\xi\in R^d\backslash\{0\}$. This is clear thanks to the positive definiteness of $D(u)$.  \\

Thanks to this proposition and the conservation-dissipative structure, we use Theorem 2.1 in \cite{YY} and get the following conclusion.

\begin{Theorem}
Under the conditions of Theorem 3.1, let $s>d/2+1$ be an integer. Assume that $\tilde U(\varepsilon)$ as initial data for PDEs \eqref{31} and $\bar u(\varepsilon)$ for \eqref{46} are in $H^s(R^d)$ for $\varepsilon>0$, satisfy
$$
\|\tilde u(\cdot,\varepsilon)-\bar u(\cdot,\varepsilon)\|_s=O(\varepsilon^2)
$$
with $\tilde u(x,\varepsilon)$ the conserved mode of $\tilde U(x,\varepsilon)$, and
all the components corresponding to densities have positive lower bounds. Then there exist $\varepsilon$-independent positive constants $T_*>0$ and $K(T_*)$ such that the solutions to PDEs \eqref{31} and \eqref{46} with the above initial data, denoted by $U^\varepsilon(x,t)$ and $u_\varepsilon(x,t)$, are in $C([0,T_*],H^s(R^d))$ and
$$
\sup_{t\in[0,T_*]}\|u^\varepsilon(\cdot,t)-u_\varepsilon(\cdot,t)\|_s\leq K(T_*)\varepsilon^2,
$$
where $u^\varepsilon(x,t)\equiv (\rho^\varepsilon,\rho^\varepsilon
\mathbf V^\varepsilon,\rho_1^\varepsilon,\cdots,\rho_{N-1}^\varepsilon)$ is the conserved mode of $U^\varepsilon(x,t)$.
\end{Theorem}

The notation used in Theorem 4.1 is standard: For a nonnegative integer s, $H^s(R^d)$ is the space of functions whose distribution derivatives
of order $\leq s$ are all in $L^2$ and we use $\|U\|_s$ to denote the standard norm of $U \in H^s$. $C([0, T ],X)$ represents the space of continuous functions on $[0, T ]$ with values in a Banach space $X$.

Finally, we give a detailed comparison with the multicomponent diffusion law derived in \cite{Lam}.

\begin{rem}
In \cite{Lam}, Lam introduced $N$ numbers $\omega_i$ satisfying $\sum\limits_{i=1}^N\omega_i\neq 0$ and replaced the collision coefficients $K_{ij}$ with $\hat K_{ij}\equiv K_{ij}+\omega_i\rho_j$ based on the zero-net-flux condition. By using the invertibility of the rank-one modification $[\hat K_{ij}]$ of the singular matrix $[K_{ij}]$, he derived the following multicomponent diffusion law
$$
\mathbf J_i=-\varepsilon\sum\limits_{j=1}^N\rho_i\bar D_{ij}\bar {\mathbf d}_j, \quad i=1, 2, \cdots, N,
$$
where $[\bar D_{ij}]=p[\hat K_{ij}]^{-1}$ and $\bar {\mathbf d}_j=\nabla \dfrac{p_j}{p}+(\dfrac{p_j}{p}-\dfrac{\rho_j}{\rho})\nabla(\ln p)$ with $p=\sum\limits_{l=1}^Np_i$.
It is not difficult to verify the following relation
$$
p(\dfrac{\bar{\mathbf d}_j}{\rho_j}-\dfrac{\bar{\mathbf d}_N}{\rho_N})=\dfrac{\nabla p_j}{\rho_j}-\dfrac{\nabla p_N}{\rho_N}
$$
between our thermodynamic forces and Lam's.
Obviously, Lam's multicomponent diffusion law is not completely determined in general, for the diffusion matrix $[\bar D_{ij}]$ depends on the arbitrary parameters $\omega_i$. Moreover, it is not clear whether $[\bar D_{ij}]$ is symmetric or positive-definite, while so is ours $[D_{ij}(u)]$.

\end{rem}

\section*{Acknowledgments}

The authors are grateful to Professor Sau-Hai (Harvey) Lam for valuable discussions. This work was supported by the Tsinghua University Initiative Scientific Research Program (grants 20121087902 and 20131089184) and by the National Natural Science Foundation of China (NSFC 11471185).

\end{spacing}

\end{document}